\documentstyle[psfig]{elsart}    
\begin{document}
\begin{frontmatter}
\title{An Energy Feedback System for the MIT/Bates Linear Accelerator}
\author[mit]{D.H. Barkhuff}, 
\author[lou]{S.P. Wells}, 
\author[cal]{T. Averett}, 
\author[uoi]{D.H. Beck}, 
\author[umd]{E.J. Beise}, 
\author[mit]{D. Cheever}, 
\author[mit]{G. Dodson}, 
\author[mit]{S. Kowalski}, 
\author[cal]{R.D. McKeown},
\author[cal]{B.A. Mueller},
\author[vpi]{M. Pitt}, 
\author[umd]{D. Spayde}, 
\author[mit]{C.~Tschal\ae r},
\author[mit]{A. Zolfaghari}
\address[mit]{Bates Linear Accelerator Center, 
              Laboratory for Nuclear Science and Department of Physics,
              Massachusetts Institute of Technology, 
              Cambridge, Massachusetts 02139}
\address[lou]{Center for Applied Physics Studies, 
              Louisiana Tech University, 
              Ruston, Louisiana 71272}
\address[cal]{Kellogg Radiation Laboratory, 
              California Institute of Technology, 
              Pasadena, California 91125}
\address[uoi]{University of Illinois at Urbana-Champagne, 
              Urbana, Illinois 61801}
\address[umd]{University of Maryland, 
              College Park, Maryland 20742}
\address[vpi]{Virginia Polytechnic Institute and State University, 
              Blacksburg, Virginia 24061-0435}
\begin{abstract}
We report the development and implementation of an energy feedback system for
the MIT/Bates Linear Accelerator Center. General requirements of the system are
described, as are the specific requirements, features, and components of the
system unique to its implementation at the Bates laboratory. We demonstrate that
with the system in operation, energy fluctuations correlated with the 60~Hz line
voltage and with drifts of thermal origin are reduced by an order of magnitude. 
\end{abstract}
\end{frontmatter}
29.17.+w, 29.27.Eg, 29.27.-a, 29.50.+v

\section{Introduction and Motivation}

Beam energy stability is of fundamental importance in any scattering experiment.
Not only are the results of such experiments sensitive to fluctuations in beam
energy, but energy variations can significantly affect transmission through beam
line elements that transport the beam to the experimental area. Instability can
therefore result in beam losses, the subsequent creation of large backgrounds in
the experimental detectors, and, especially if reliable extraction of
observables depends critically on proper background subtraction, an increase in
systematic uncertainties. In addition, beams are often injected into internal
storage rings, where mismatch between the injected beam energy and the ring
energy can again result in the significant losses and large backgrounds
associated with beam scrape-off. It is also true that the energy of pulsed
electron beams can be susceptible to slow drifts, and, especially if DC supply
voltages and power are coupled to the AC line, to fluctuations at 60~Hz. 

These issues are important at the MIT/Bates Linear Accelerator Center, where
pulsed electron beams with energies of up to 1~GeV are generated for transport
to two main experimental areas and for injection into the new South Hall Ring
(SHR). In order to improve beam energy stability, we have recently designed and
implemented a feedback system capable of adjusting the beam energy in response
to both 60~Hz fluctuations and slow drifts. A beam position monitor (BPM)
installed in a dispersive region of the beam line allows the energy of each
pulse to be measured. The BPM signal, which is sensitive to changes in the
amplitude or phase of any of the twelve klystrons that supply radio frequency
power for the accelerator, is the system's single ``dial''. A phase shifter, the
system's single ``knob,'' is installed on one of the klystrons to allow small,
rapid, computer controlled energy adjustments to effectively compensate 
for those changes. 

This work was primarily motivated by a need to minimize beam energy variations
on both short and long time scales for two experimental initiatives at
MIT/Bates. First, the SAMPLE experiment requires a beam of exceptional stability
to measure the parity violating spin-dependent cross section asymmetry of only a
few parts per million in elastic electron scattering at backward angles from
unpolarized hydrogen and deuterium targets[1].
During test runs, it
was found that uncontrolled beam energy fluctuations led to scrape-off on a pair
of energy limiting slits upstream of the SAMPLE apparatus. Sensitivity to the
small parity violating observable was reduced by the background
subsequently created in the detectors. Second, the maintenance of the beam's
energy and position within narrow limits is a prerequisite for the efficient
operation of the SHR, a vital component of a new program of internal target
experiments with the Bates Large Acceptance Spectrometer Toroid
(BLAST). 

The discussion is divided into four parts. In Section 2, we describe particular
features of the accelerator and the beam at MIT/Bates that had significant
impact on the design of the feedback system. We discuss in Section 3 the
components of the feedback system and their functions in detail. In Section 4,
we present results that demonstrate the marked improvement in beam energy
stability with the energy feedback system. Finally, in Section 5, we summarize
our most significant results and conclusions. 

\section{Beam Structure and Beam Energy Instability}

At Bates, beams of electrons are accelerated with longitudinal electric fields
oscillating in a series of accelerating cavities at a radio frequency (RF) of
2856~MHz. RF power is delivered to the cavities through wave guides
from up to twelve klystrons, and beams of different energies are prepared by
adjusting the RF amplitude in each klystron. Electrons are injected at all RF
phases, but those injected outside of a $120^{\circ}$ phase window are
immediately ``chopped'' or deflected into a metal beam stop by a transverse RF electric
field. The remaining electrons are ``bunched,'' or compressed with a
longitudinal RF electric field into a phase window of about $2^{\circ}$. The RF
phases of all but one klystron are optimized with respect to a common
reference signal so that the crest of the RF field in each accelerating cavity
coincides with the $2^{\circ}$ beam ``bunches'' as they pass through it. 
Klystron 6B, designated the ``vernier'', is shifted away from its crest. The
phase of the vernier can be adjusted, compensating for drifts in the RF phase or
amplitude of the other klystrons. 

The RF transmitters deliver power only in short bursts followed by a substantial
recovery period. A duty cycle, about 1\% for the Bates facility, is therefore
superimposed on the RF microstructure of the beam. Typically, the accelerator is
configured for beam pulses of 3-25~$\mu$s in duration at a rate of 600~Hz. This
time structure matches the frequency with which the beam polarization can be
reversed at the Polarized Electron Source (PES). At the PES, electrons are
photoemitted from a GaAs crystal illuminated by circularly polarized laser light
with a helicity that can be chosen randomly at 600~Hz. 

\begin{figure}
\begin{center}
\vspace{20mm}
\centerline{\psfig{figure=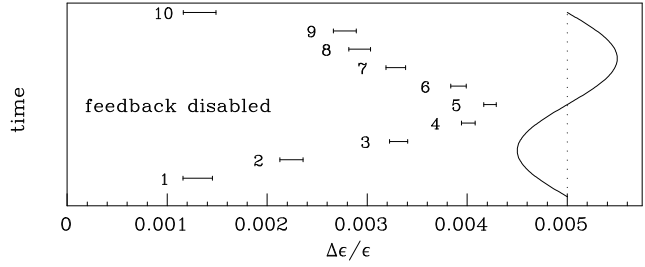,width=3.5in}}
\caption{Fractional energy change as a function of time slot over the range of 
the 60~Hz AC power cycle. Without feedback, beam energy varies by 0.3\% over the 
power cycle, superimposed to set the vertical time scale.}
\label{fig:60hz1}
\end{center}
\end{figure}

The beam pulses are synchronized with the laboratory's 60~Hz AC line voltage, so
that every pulse is associated with one of ten ``time slots,'' $1\le n\le 10$,
each of which has a unique phase angle ${{n\pi}\over{5}}$ with respect to the
60~Hz AC power cycle[2].
The first time slot is triggered by the
positive-going zero crossing of the AC line voltage. 
Ideally, the properties of the beam
would be independent of time slot.
However, because DC power to the RF transmitters and the magnetic
beam line elements is not perfectly isolated from the AC power, beam properties
can fluctuate in ways that are highly correlated with the 60~Hz AC line, and
therefore with time slot. For example, the strong dependence of beam energy on
time slot is shown in Fig. \ref{fig:60hz1}. These data, obtained over a period
of about 10~s with the energy feedback system disabled, show the beam's
fractional deviation $\Delta\epsilon /\epsilon$ from the nominal beam energy for
each time slot. The 60~Hz AC line voltage is superimposed to show the relative
timing of each pulse and to set the vertical time scale. This figure
demonstrates that beam energy variation in a single time slot can be at least an
order of magnitude smaller than its variation over an entire power cycle.
Overall variation can therefore be reduced significantly by a system capable of
applying rapid time slot dependent energy corrections. 

\begin{figure}
\begin{center}
\vspace{30mm}
\centerline{\psfig{figure=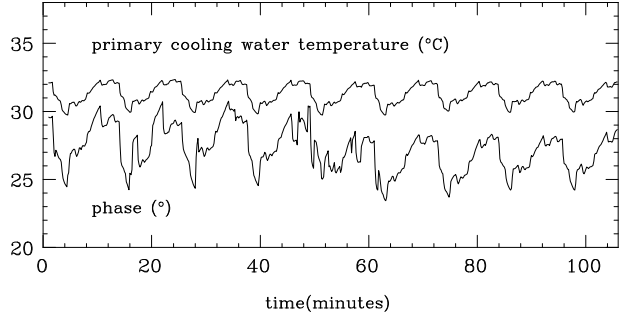,width=3.5in}}
\caption{Temperature variations in the laboratory cooling water, in $^{\circ}$C,
as a function of time. Also plotted is the RF phase of the electric field, in
$^{\circ}$, in one of the accelerating cavities.} 
\label{fig:phase}
\end{center}
\end{figure}

Beam energy is also subject to fluctuations at other frequencies, particularly
slow drifts with characteristic times between 10~s and 1000~s. In Fig.
\ref{fig:phase}, for example, the RF phase of the electric field in one of the
klystrons with respect to the common RF reference is shown to correlate with
temperature variations in the water that provides cooling to the transmitters
and magnets. Uncompensated RF phase variations can lead to beam energy 
variations that, in turn, often result in unacceptable background levels.

Our feedback system is designed to compensate for both types of beam energy
instability. Required of the system is the ability to 
\begin{itemize}
\item 
monitor the beam energy $\epsilon _n$ in each time slot $n$, and accurately
determine its average over a time interval $\Delta t$ which is small compared
with the characteristic period of slow drifts; 
\item 
estimate, for each time slot, the difference $\Delta \epsilon _n$ between the
beam energy $\epsilon _n$ and some ideal energy $\epsilon_0$;
\item 
estimate and store, for each time slot, the phase shift 
$\Delta \phi _n \equiv -\Delta \epsilon _n 
({{\partial \epsilon}\over{\partial \phi}})^{-1}$ required
to minimize $\vert \Delta \epsilon _n \vert$; 
\item 
shift the phase of the energy correction klystron sufficiently in advance of
each time slot so that the phase is stable before the beam is injected. 
\end{itemize}

\section{Instrumentation and Operation}

\begin{figure}
\begin{center}
\vspace{29mm}
\centerline{\psfig{figure=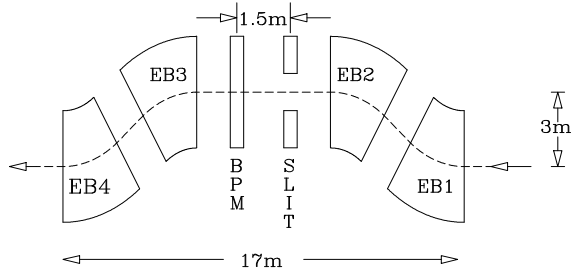,width=3.5in}}
\caption{A plan view of the chicane, showing the BPM, the energy limiting slits, 
and the four dipole magnets EB1, EB2, EB3 and EB4. The central ray is represented as a 
dashed line.}
\label{fig:dipole}
\end{center}
\end{figure}

There are three main components of the energy feedback system. First, energy is
monitored with a BPM installed between the two dipole pairs of a magnetic
chicane located downstream of the accelerator and shown schematically in plan
view in Fig. \ref{fig:dipole}. The chicane disperses the beam horizontally in
the region between the dipole pairs. By definition, the trajectory of electrons
with the ``central'' energy passes equidistant from a pair of moveable, heavy metal,
water cooled slits, positioned to the left and right of the center of the beam
line. Electrons of higher energy and higher rigidity follow shorter trajectories
than electrons of the central energy and therefore pass through the 
chicane to beam left. In contrast, lower energy electrons follow longer
trajectories that pass through the chicane to beam right. Typically, the slits are
positioned symmetrically with a separation of 33~mm, limiting the
beam energy spread in the chicane to $\pm0.5\%$[3].

The BPM, located 1.5~m downstream of the slit, is a non-resonant RF cavity
perpendicular to the beam with a diode at each end. The RF pulse structure of
the beam induces oscillation in both diodes, with a relative RF phase
proportional to the beam's displacement from the center of the cavity. Due to
the correlation of 33~mm/\% between horizontal beam position and energy, the
phase is also proportional to the beam energy's relative deviation from the
central energy. The error signal, a voltage proportional to this phase, is
produced by the BPM's output stage, and is integrated over the duration of each
pulse and digitized in a 16 bit ADC. The intrinsic position resolution of this
BPM is of order 50~$\mu$m, with typical output voltages, before amplification,
of about 3~mV/mm displacement.

The second component of the feedback system is a remotely controlled ferrite
core phase shifter with a 12 bit digital interface[4].
This digital
phase shifter (DPS) is installed on the vernier klystron. Changes in the RF
phase stabilize in about 1~ms and can be made in increments as small as 3~mr. 

\begin{figure}
\begin{center}
\vspace{24mm}
\centerline{\psfig{figure=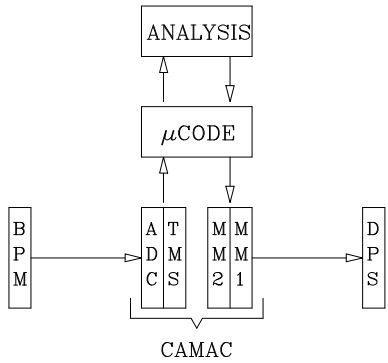,width=3.5in}}
\caption{An overview of the energy feedback system. A BPM signal is digitized in
a CAMAC ADC and read after each beam burst. A time slot scaler (TMS),
incremented at 600Hz and cleared on the positive-going zero crossing of the 60
Hz power cycle, identifies the time slot of every event in the data stream. A
low-level microcode controls data acquisition from CAMAC, data shipment to a
high-level analyzer, and the download of bit patterns to the digital phase
shifter (DPS). The analyzer computes the appropriate phase shift and bit pattern
for each time slot.} 
\label{fig:schematic}
\end{center}
\end{figure}

The third component is a computer controlled interface between the error signal
and the phase shifter. The interface performs three functions. Two of the three
functions, data acquisition and energy correction, are controlled by a low level
microcode executing CAMAC read and write instructions in synchronization with
the 600~Hz pulse rate. Data analysis, a third function of the interface, is
performed asynchronously by a separate program. The flow of information between
the components of the energy feedback system is indicated schematically by the
diagram in Fig. \ref{fig:schematic} 

Before enabling the feedback system, the accelerator is prepared according to a
standard procedure. First, the ten digital bit patterns that encode the phase
shift for each time slot are initialized to the middle of their full range, which
is limited in software to 45$^{\circ}$. Next, all klystrons are phased with
respect to the common RF reference. To first order, the beam energy is then
independent of small drifts in phase from the RF crest. The phase of the vernier
klystron is then manually shifted away from its crest by about 22$^{\circ}$,
half of the system's full range. In this configuration, beam energy is quite
sensitive to automatic adjustments by the feedback system of the vernier's RF
phase.

Occasionally, the beam operators adjust one or more of three software parameters in
order to optimize the performance of the system. One of these parameters is the
ideal energy $\epsilon_0$, toward which the feedback system is programmed to
drive the actual beam energy. This ideal energy is chosen to optimize the
experimental running conditions, and must be within 0.5\% of the
central energy. Otherwise, the beam will collide with the energy limiting 
slits. The operators can also select the number of beam pulses to be sampled before a new
set of phase shifts is determined. This affects the speed with which the system
corrects the energy, and the statistical precision of the corrections. A third
parameter specifies the sign and magnitude of the phase shifts with respect to
beam energy deviations. The value of this parameter, used to control overshoot and
undershoot, has normally been obtained from an experimental measurement of
${{\partial \epsilon}\over{\partial \phi}}$ and has been found to be near 
0.0043\%/mr (0.075\%/degree). However, its optimum value has been shown to vary
by at most 10\% for widely varying beam tunes and is now rarely remeasured.
Following these preparations, energy feedback is enabled. 

At 1.6~ms prior to each pulse, the CAMAC system extracts a digital bit pattern
corresponding to the phase shift for the next time slot from one of ten
locations in a LeCroy 8206A CAMAC memory module (MM1) and transfers the pattern
to a TTL output register (DSP PR-612) connected to the phase shifter's digital
interface. Within 1~ms, the phase shift of the vernier klystron is stable. After
each pulse, the CAMAC system reads the digitized BPM error signal along with a
label identifying the current time slot. The digitized data are packaged and
distributed to the analysis program at 7.5~Hz. 

For each time slot $n$, the analyzer computes an average error $\Delta
\epsilon_n$ and a phase shift $\Delta \phi_n$ from a total accumulation of about
1000 beam pulses (100 in each time slot). These new phase shifts are encoded as
digital bit patterns and downloaded to the CAMAC system. However, because the
analyzer operates asynchronously, a download of the bit patterns directly into
the memory module MM1 can interrupt the 600~Hz access of the microcode to its contents. To
prevent this, the new bit patterns are written asynchronously into ten locations
of a separate memory module MM2. When all ten patterns have been loaded, an
additional flag is set to indicate a Data Ready condition. 
Prior to the first beam pulse in a ten time slot sequence, the Data Ready condition triggers the
transfer to MM1 of the new bit patterns in MM2. The transfer is controlled by
the microcode, requires 400~$\mu$s, and is completed 1~ms before the next memory
access. Although the system's frequency response would normally be limited to
about 1~Hz by the time required to acquire a statistically significant sample,
it effectively compensates for 60~Hz fluctuations by simultaneously implementing
an independent feedback loop for each of the ten time slots. 

Separate analog signals, corresponding to the highest and lowest of the ten
digital phase shift values for a given sample cycle, are piped from the CAMAC
acquisition hardware to the main accelerator control room and monitored as a
function of time. Should slow but constant energy drifts cause the feedback
algorithm to approach either its high or low software limit, the beam can be
rephased, and the digital bit patterns can be reset for all time slots to the
center of the feedback system's software range. 

\section{Energy Feedback Performance}
\begin{figure}
\begin{center}
\vspace{35mm}
\centerline{\psfig{figure=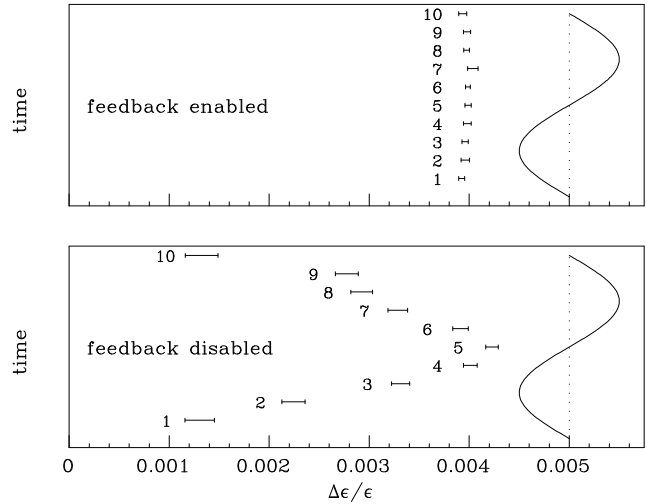,width=3.5in}}
\caption{Fractional energy change as a function of time slot over the range of 
the 60~Hz AC power cycle. The lower (upper)
panel shows the behavior of the beam with the energy feedback system disabled
(enabled). The 60~Hz AC line voltage is superimposed to set the time scale.}
\label{fig:60hz2}
\end{center}
\end{figure}

\begin{figure}
\begin{center}
\vspace{35mm}
\centerline{\psfig{figure=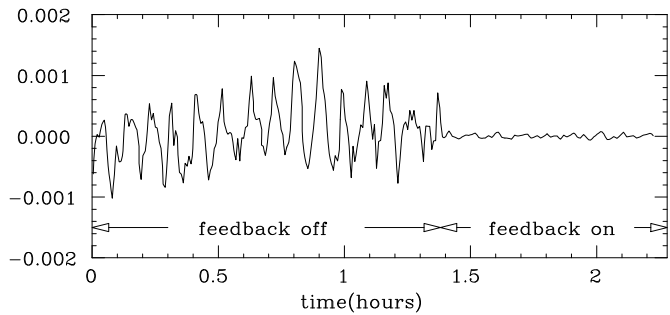,width=3.5in}}
\caption{Fractional energy change as a function of time for a single time slot,
with and without the energy feedback system.}
\label{fig:feedback}
\end{center}
\end{figure}

Two figures demonstrate the ability of the system to meet its design goals. The
first, Fig. \ref{fig:60hz2}, shows the behavior of the beam on a time scale of a
few seconds as a function of time slot, both with and without the energy
feedback system. In each of the two panels the data are averaged over an
interval of about 10~s. With feedback disabled, beam energy fluctuates over the
range of the AC power cycle by 0.3\% (lower panel). With feedback enabled
and, in this example, adjusted for an ideal energy 0.4\% above the central
energy, fluctuations are controlled at the level of about 0.02\% (lower panel).
Moreover, RMS fluctuations per time slot are reduced by a factor of 2.
The second, Fig. \ref{fig:feedback}, shows the effect of the system over long
periods of time for an individual time slot. With feedback disabled, the
characteristic magnitude of slow beam energy drifts is about 0.2\% of the
central energy. However, the feedback system, enabled in this case for a set point
equal to the central energy, effectively eliminates these drifts. 

\section{Summary and Conclusion}

Although 60~Hz AC line voltage fluctuations and slow, temperature dependent
changes in accelerator hardware can induce energy instabilities, we have
developed a reliable feedback system at the MIT/Bates Linear Accelerator which
can compensate for these changes. Before the installation of this system, 60~Hz
line voltage fluctuations induced energy fluctuations of up to 0.4\%, and slow
phase variations of thermal origin induced energy fluctuations of up to 0.2\%.
With the energy feedback system enabled, beam energy is measured on a pulse
by pulse basis while rapid changes are made in the RF phase in one of the
accelerating cavities, controlling energy fluctuations at the level of 0.02\%.

Two important sources of energy instability have been effectively eliminated,
resulting in a beam with an energy spread that is limited only by the width of
the energy distribution within a 3-25~$\mu$s beam pulse. This system
provides improved beam stability, a decrease in background due to beam
scrape-off during transport, and a simplification in the operation of the
accelerator. The energy feedback system developed here has proven to be so
successful that its use is now a routine part of the standard operating
procedure for the MIT/Bates Linear Accelerator. 

{\bf Acknowledgements}
We gratefully acknowledge the help and support of the MIT/Bates electrical
engineering and RF groups. We also thank the Rensselaer Polytechnic Institute
group, the SAMPLE collaboration, and all others who participated in several
development runs. This work was supported in part by the National Science
Foundation and in part by the Department of Energy under Cooperative Agreement
No. DE-FC02-94ER40818.A000.


\begin{thebibliography}{4}
\bibitem{ecs} {All energy deviations quoted in this article are with respect to
the BPM in the chicane. An independent Energy Compression System, located
downstream of the chicane, narrows the beam energy distribution by an additional
order of magnitude.}
\end{thebibliography}
\end{document}